\documentclass[conference]{IEEEtran}
\IEEEoverridecommandlockouts

\usepackage{algorithm}
\usepackage{algpseudocode}
\usepackage{cite}
\usepackage{subcaption}
\usepackage{amsmath,amssymb,amsfonts}
\usepackage{graphicx}
\usepackage{textcomp}
            
\def\BibTeX{{\rm B\kern-.05em{\sc i\kern-.025em b}\kern-.08em
    T\kern-.1667em\lower.7ex\hbox{E}\kern-.125emX}}
\begin{document}

\title{\huge{Adaptive Beam-Frequency Allocation Algorithm with Position Uncertainty for Millimeter-Wave MIMO Systems}
	{\footnotesize }
}

\author{
	\IEEEauthorblockN{Rafail Ismayilov\IEEEauthorrefmark{1}, Megumi Kaneko\IEEEauthorrefmark{2}, Takefumi Hiraguri\IEEEauthorrefmark{3} and Kentaro Nishimori\IEEEauthorrefmark{4}}
	\IEEEauthorblockA{\IEEEauthorrefmark{1}University of Goettingen, Goettingen, Germany}
	\IEEEauthorblockA{\IEEEauthorrefmark{2}National Institute of Informatics, Tokyo, Japan}
	\IEEEauthorblockA{\IEEEauthorrefmark{3}Nippon Institute of Technology, Saitama, Japan \hspace{5pt}
	\IEEEauthorblockA{\IEEEauthorrefmark{4}Niigata University, Niigata, Japan
		\\ e-mail: rafail.ismayilov@stud.uni-goettingen.de, megkaneko@nii.ac.jp, hira@nit.ac.jp, nishimori@ie.niigata-u.ac.jp}}
}

\maketitle
\begin{abstract}
	Envisioned for fifth generation (5G) systems, millimeter-wave (mmWave) communications are under very active research worldwide. Although pencil beams with accurate beamtracking may boost the throughput of mmWave systems, this poses great challenges in the design of radio resource allocation for highly mobile users. In this paper, we propose a joint adaptive beam-frequency allocation algorithm that takes into account the position uncertainty inherent to high mobility and/or unstable users as, e.g., Unmanned Aerial Vehicles (UAV), for whom this is a major problem. Our proposed method provides an optimized beamwidth selection under quality of service (QoS) requirements for maximizing system proportional fairness, under user position uncertainty. The rationale of our scheme is to adapt the beamwidth such that the best trade-off among system performance (narrower beam) and robustness to uncertainty (wider beam) is achieved. Simulation results show that the proposed method largely enhances the system performance compared to reference algorithms, by an appropriate adaptation of the mmWave beamwidths, even under severe uncertainties and imperfect channel state information (CSIs). \footnote{This work is supported by the Grants-in-Aid for Scientific Research (Kakenhi) nos. 17H01738 and 17K06453 from the Ministry of Education, Science, Sports, and Culture of Japan.}  
\end{abstract}

\vspace{5pt}
\section{Introduction}\label{s1}
To support high data rate requirements, mmWave-based communications are being actively considered for the future 5G systems \cite{b1}. MmWave technologies will be one of the key solutions against the severe spectrum deficiency problems of current wireless communication systems. Indeed, they would provide ultra-wide GHz spectrum usage at higher frequency bands, ranging from 30 to 300GHz, and in particular in the 28, 38, 60 GHz and E-bands (71-76 and 81-86GHZ), creating multi-Gbps data throughput. 

Enabling mmWave technology imposes great challenges in the design of PHY and MAC layers \cite{b2}. Until now, most of the works have dealt with the PHY aspects such as developing multiple-input multiple-output (MIMO) and beamforming techniques for mmWave wireless systems. To alleviate the system complexity inherent to mmWave massive MIMO, a promising solution lies in the concept of hybrid beamforming (HBF), which uses a combination of analog beamformers in the radio frequency (RF) domain, together with digital beamforming in the baseband \cite{b3}. However, effective radio resource allocation (RRA) and interference mitigation methods for mmWave multi-user systems are still in the early research phase. 

\begin{figure}
	\centering
	\includegraphics{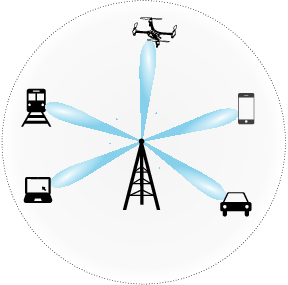}
	\caption{5G mmWave system with multi-user beamforming}
	\vspace{-0.5em}
	\label{fig1}
\end{figure}

In \cite{b5}, an RRA method for a multi-beam multi-user mmWave system was proposed. To improve spectral efficiency, an interference management and scheduling scheme based on coordinated multipoint transmission (CoMP) is designed. However, such a coordination scheme entails significant burden in terms of signaling overhead for control information sharing and channel state information (CSI) feedbacks, one of the major drawbacks of mmWave massive MIMO systems pointed out in \cite{b3}.

Authors in \cite{b6} proposed a multiuser beam-frequency scheduling method for mmWave system with analog beamformers. Each user feeds back its CSI which is assumed to be perfectly known at the base station (BS). However, such an assumption is not suited for a system with mobile user equipments (UE). In addition, interference among allocated beams were not considered.

In \cite{b4}, a beam-frequency allocation algorithm for throughput maximization under user quality of service (QoS) constraints was considered. Throughput optimization problem is considered under user QoS requirements, while dealing with mainlobe interferences. However, the proposed algorithm is designed for fixed beamwidths, and also considers perfect CSI knowledge and fixed user positions without uncertainties, which are impractical assumptions for high mobility users such as unmanned aerial vehicle (UAV) networks which are under strong research focus. In particular, estimated positions are inaccurate in UAV systems, due to imperfect sensor calibration, or environmental conditions such as wind speed and direction. Another important aspect is the update rate of position information. For instance, the update rate of GPS positioning information is typically every 0.1-1 sec. \cite{b8}, which is larger than the scheduling frame length in target wireless systems. This signifies that the estimated position becomes rapidly obsolete. Therefore, taking into account uncertainties in terms of position and CSI is crucial in the design of efficient and realistic RRA and interference management, especially in mmWave MIMO systems where accurate pencil beams are essential \cite{b7}.

In this work, we propose a joint adaptive beam-frequency allocation method for multi-user mmWave systems under user position uncertainties and imperfect CSI knowledge at the BS. We design a fairness-aware scheme that adapts the allocated beamwidth depending on the individual user QoS requirements, and user position uncertainty levels. One of the key aspects of our scheme is to adaptively optimize the allocated beamwidth/amount of subbands such that the best trade-off between system performance and robustness towards user channel/position uncertainties is achieved. The simulation results show that the proposed scheme largely outperforms benchmark beam-frequency allocation schemes in terms of system proportional fairness, even under severe uncertainty conditions.

\section{System Model}\label{s2}

\subsection{System and Metrics}\label{ss2_A}

We assume a single mmWave BS with HBF equipped with $M$ analog beamformers and serving $K$ users as in \cite{b4}. Each analog beamformer covers a $\frac{2\pi}{M}$ sector of the cell, hence $M$ parallel beams may be simultaneously transmitted (i.e., one beam per beamsector) to multiple UEs in each scheduling time frame. Frequency channels are divided into $N$ subbands, each of bandwidth $W$. Thus, in total, there are $MN$ resource blocks per frame. Additionally, we assume that each subband may be allocated to only one user at a time. For simplicity, we suppose that Tx and Rx beamwidths $\theta_t$ and $\theta_r$ as well as their directivity beam gains are equal,  i.e., $\theta_t = \theta_r = \theta$. Tx/Rx directional beam gains are approximated following \cite{b7},
\begin{equation}\label{eq1}
	\left\{\begin{array}{ll}
	g^{tm} = g^{rm} = \frac{2\pi-(2\pi-\theta)\epsilon}{\theta} &, \text{in Tx/Rx mainlobe},\\ 
	g^{ts} = g^{rs} = \epsilon &, \text{in Tx/Rx sidelobe},
	\end{array}\right.
\end{equation}
where $\epsilon \ll  1$. Mainlobe interference occurs when the BS allocates the same subband to multiple UEs in the same beamsector. By contrast, sidelobe interference occurs among beams from different beamsectors but in the same subband. Unlike in \cite{b4}, our beam-frequency allocation method will ensure that each subband will be allocated to a unique user in each beam/beamsector, thereby eliminating any mainlobe interference. Thus, only sidelobe interference will occur. The resource allocation decision is represented by $\Phi = \begin{Bmatrix} \phi_{k,m,n}|1\le k\le K, 1\le m\le M,1\le n\le N \end{Bmatrix}$, where $\phi_{k,m,n} \in\begin{Bmatrix} 0,1 \end{Bmatrix}$ indicates whether user $k$ is assigned subband $n$ on beam $m$. The signal-to-interference-plus-noise ratio (SINR) for user $k$ at beam $m$ on subband $n$ is given by
\begin{equation}\label{eq2}
	\gamma_{k,m,n} = \frac{P_{k,m,n}  g_k^{tm}   g_k^{rm} |h_{k,m,n}|^2 d_k^{-\alpha}}{\displaystyle\sum_{j \ne m}^{M}\displaystyle\sum_{i \ne k}^{K}P_{i,j,n} |h_{k,m,n}|^2 d_k^{-\alpha}  g_i^{ts} g_k^{rs}  \phi_{i,j,n} + N_0  W}, 
\end{equation}
where $P_{k,m,n}$ is the BS transmit power, $g_k^{tm}$ and $g_k^{rm}$ are the Tx and Rx beam gains for user $k$, and $h_{k,m,n}$ is the channel coefficient between the BS and user $k$ on beam $m$ and subband $n$, assumed to follow Rayleigh fading, i.e., $h_{k,m,n} \sim  \mathcal{CN}(0,1)$. The distance from the BS to user $k$ and the pathloss exponent are denoted by $d_k$ and $\alpha$, respectively. The first term in the denominator expresses the total sidelobe interference, i.e., sum over signals towards other users $i$ on other beams $j$ but on the same subband $n$. Note that since no users share the same subband in the same beamsector, we have $g_{i}^{ts}g_{k}^{ts}=\epsilon^2$ from \eqref{eq1}. $N_0$ is the power spectral density of the additive white Gaussian noise (AWGN). From \eqref{eq2}, the achievable data rate for user $k$ can be obtained by
\begin{equation}\label{eq3}
	R_k = \sum_{m=1}^{M}\sum_{n=1}^{N}W\log_2(1+\gamma_{k,m,n})\phi_{k,m,n}.
\end{equation}
\begin{figure}
	\centering
	\includegraphics[width=0.99\linewidth]{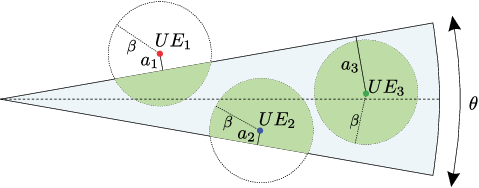}
	\caption{Illustration of the user position uncertainty model}
	\vspace{-0.5em}
	\label{fig2}
\end{figure}
Most previous works such as \cite{b4}\hspace{1pt}\cite{b6} assume perfect instantaneous CSI knowledge for all users, on each subband and beam. However, this causes tremendous CSI feedback overhead. In addition, with mobile users under high position/channel uncertainty, such CSIs become quickly obsolete. Therefore, in this work, we assume that each user only reports its position and large-scale channel fading conditions to the BS in every time frame, thereby greatly decreasing the amount of CSI feedback. Thus, the proposed allocation will be optimized based on position/distance knowledge only, using the imperfect SINR
\begin{equation}\label{eq22}
	{\gamma}'_{k,m,n} = \frac{P_{k,m,n}  g_k^{tm}   g_k^{rm} d_k^{-\alpha}}{\displaystyle\sum_{j \ne m}^{M}\displaystyle\sum_{i \ne k}^{K}P_{i,j,n} d_k^{-\alpha}  g_i^{ts} g_k^{rs} + N_0  W},
\end{equation}
where we have made the reasonable assumption that all subbands are occupied, i.e., $\phi_{i,j,n}=1$, and assuming equal power allocation. However, the actual perceived rates are determined based on real SINRs \eqref{eq2} after allocation. 
\subsection{Position Uncertainty Model}\label{ss2_B}
We build our position uncertainty model based on estimated and actual user positions. The level of user position uncertainty is expressed by a parameter $\beta$ in meters. The estimated position is reported to the BS which uses it for resource allocation. However, the actual position will be assumed to be uniformly distributed within a circle centered at the estimated position, with radius $\beta$. In the case of perfect position knowledge, we have $\beta = 0$.

In Fig. \ref{fig2}, the BS receives the estimated position information from UE$_1$, UE$_2$ and UE$_3$, marked with red, blue and green dots respectively. In addition, $a_1$, $a_2$ and $a_3$ denote the distance from the estimated position to the closest beam edge, for each UE respectively. From Fig. \ref{fig2} we can see that UE$_2$ and UE$_3$'s estimated positions are inside the beam coverage of beamwidth $\theta$. Furthermore, UE$_3$'s actual position will be always covered by the operating beam. 
\section{Problem Formulation}\label{s3}
From \eqref{eq1} we can observe that, beam gains in the mainlobe are increasing with smaller beamwidth $\theta$, which subsequently increases SINR and data rate. While smaller beamwidths increase throughput, they greatly reduce the number of users within beam coverage. This may largely degrade the system fairness. Therefore, the beamwidth should be adapted in order to achieve a good throughput/fairness trade-off. In addition, we consider the effect of user position uncertainty in this problem, for which small beamwidths may lead to significant performance degradation, since the actual UE position will have a higher probability to be outside the allocated beam coverage. We consider the problem of beam-frequency allocation under user position uncertainties and QoS requirements. Our goal is to optimize proportional fairness by jointly considering the beam direction/width and subband allocation, where the beamwidth can be adapted in each frame and sector, according to uncertainty levels.
\begin{figure}
	\begin{subfigure}{.25\textwidth}
		\centering
		\includegraphics[width=0.99\linewidth]{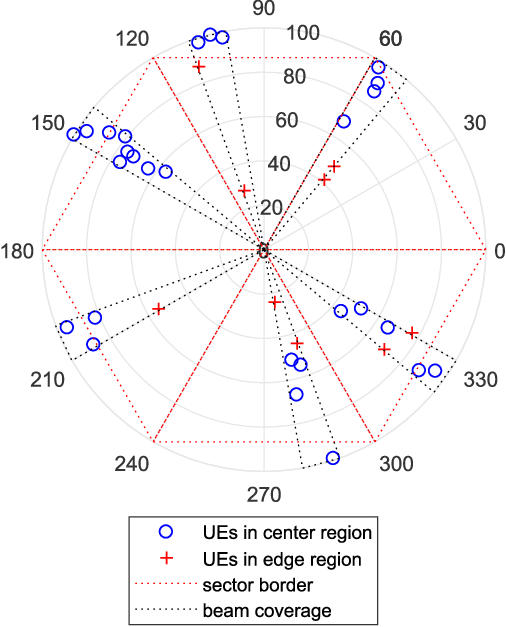}
		\caption{$\delta = 1.5m$}
		\label{sfig1}
	\end{subfigure}%
	\begin{subfigure}{.25\textwidth}
		\centering
		\includegraphics[width=0.99\linewidth]{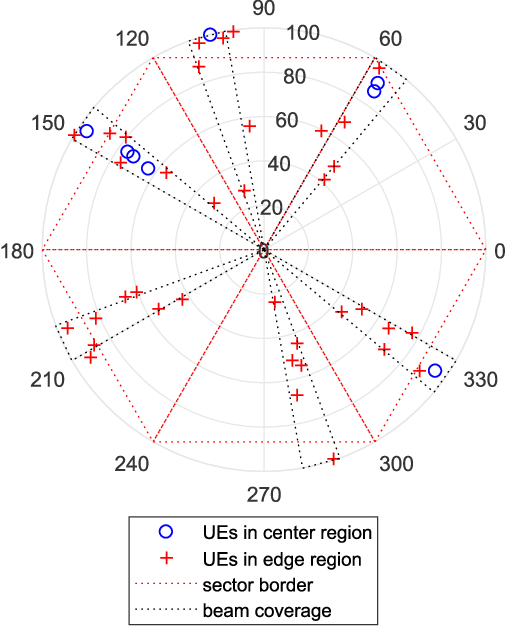}
		\caption{$\delta = 4.5m$}
		\label{sfig2}
	\end{subfigure}
	\caption{Effect of threshold ($\delta$) adjustment towards the sets of center and edge users ($\beta = 3m$ and $\theta = 10^o$)}
	\vspace{-0.5em}
	\label{fig3}
\end{figure}

In our formulation, the user position uncertainty is tackled as follows. We define a threshold $\delta$, ($0<\delta \le \beta$) and divide all UEs into two groups: center and edge UEs and denote these sets as $\mathcal{C} = \begin{Bmatrix} \text{UE}_k|a_k>\delta,1 \le k \le K \end{Bmatrix}$ and $\mathcal{E} = \begin{Bmatrix} \text{UE}_k|a_k \leq \delta,1 \le k \le K \end{Bmatrix}$, as shown in Fig. \ref{fig3}. Compared to center UEs, edge UEs have a higher probability that their actual position will get out of coverage of their allocated beam, in which case they would perceive rate $R_k=0$ and their allocated resources would be wasted. Hence, we define two different QoS requirements for each class such that $R_{min}^{\mathcal{C}} > R_{min}^{\mathcal{E}}$, where $R_{min}^{\mathcal{C}}$ and $R_{min}^{\mathcal{E}}$ indicate the minimum rate requirement for center and edge UEs, respectively. 

Thus, beamwidth $\theta$ and threshold $\delta$ are key parameters to be optimized in our proposed algorithm. Note that, for given $\theta$, by increasing $\delta$ we protect more edge UEs. Similarly for constant $\delta$, a larger $\theta$ increases the number of center UEs.

Thus, our optimization problem is formulated as
\begin{subequations} \label{eq4}
	\begin{alignat}{3}
		& \max\limits_{\theta,\delta,\Phi}
		&\quad & \Gamma = \sum_{k=1}^{K}\log(R_k) & \tag{\ref{eq4}}\\ 
		& \text{s.~t.} 
		& &  \left\{\begin{matrix} 
		R_k \ge R_{min}^{\mathcal{C}},\hspace{3pt} k \in \mathcal{C},\\ 
		R_k \ge R_{min}^{\mathcal{E}},\hspace{3pt} k \in \mathcal{E}.
		\end{matrix}\right., &\quad \forall k,  \label{eq4a}\\
		  &   &   & \sum_{k=1}^{K}\sum_{m=1}^{M}\phi_{k,m,n} \le M,                 & \forall n, \label{eq4b}                     \\
		  &   &   & \sum_{k=1}^{K}\phi_{k,m,n} \in \begin{Bmatrix}0,1\end{Bmatrix}, & \quad \quad \quad \forall m,n, \label{eq4c} \\
		  &   &   & 0 < \theta \le \frac{2\pi}{M}                                   & \quad \quad \quad \label{eq4d}              \\
		  &   &   & 0 < \delta \le \beta                                            & \quad \quad \quad \label{eq4e}              
	\end{alignat}
\end{subequations}
Constraint \eqref{eq4a} ensures that each UE in $\mathcal{C}$ or $\mathcal{E}$ receives its minimum required rate. \eqref{eq4b} represents the maximum number of beams transmitted simultaneously. \eqref{eq4c} indicates that each subband $n$ in beam $m$ is allocated to at most one UE. The operating beamwidth cannot exceed sector coverage as expressed by constraint \eqref{eq4d}. \eqref{eq4e} enforces that, threshold $\delta$ must be below the maximum uncertainty level $\beta$.

\section{Proposed Adaptive Joint Beam-Frequency Allocation Algorithm}\label{s4}
To make the problem tractable, the proposed algorithm will mainly focus on the beam frequency allocation problem in \eqref{eq4} subject to the constraints, given discrete sets of beamwidths $\theta$ and thresholds $\delta$. Then, beamwidth $\theta$ will be iteratively adapted, given threshold $\delta$. 

First, given initial $\theta$, the BS selects the beam direction that covers the highest number of users, in each sector. A key aspect of the proposed algorithm is that, it allocates resources to all users whose actual positions will be likely to be within that beam coverage. Therefore, the edge users set $\mathcal{E}$ includes not only the edge users from the selected beam, but also those in the two adjacent beams. Given these user sets, the algorithm allocates subbands for proportional fairness maximization. The concrete steps of the proposed algorithm are given in Algorithm \ref{algo1} and explained below.  

In the first step, given the reported user positions, the BS defines the set of UEs in each beam $\omega_m = \begin{Bmatrix}b_v| 1 \le v \le V \end{Bmatrix}$, where $b_v$ denotes supported beam indices and $V = \frac{2 \pi}{\theta M}$ is the maximum number of available beams in each beamsector. Collection of beamsector sets is expressed as $\Omega = \begin{Bmatrix}\omega_m|1 \le m \le M\end{Bmatrix}$. For each beamsector, the BS selects the beam $b_v$ containing the maximum number of UEs and sorts them in the order of increasing distance from the BS. If the number of UEs is identical in two or more beams within a beamsector, one beam direction is randomly selected. From the selected beam $b_v$, the BS defines its adjacent beams $b_{v+1}$ and $b_{v-1}$.
\begin{algorithm}[t]
	\caption{RRA for mobile UEs with position uncertainty}
	\begin{algorithmic}[1]
		\State \textbf{Initialization:} $\delta$
		\State Receive estimated position reports from UEs
		\State Set allocation matrix $\Phi_{\theta} \gets \varnothing$
		\For {$m=1:M$}
		\For {$\theta=\theta_{min},...,\theta_{max}$}
		\State \parbox[t]{\dimexpr\linewidth-\algorithmicindent}{Select beam $b_v^{\theta} \in \omega_m$ with maximum number \\ of UEs \strut} 
		\State Define adjacent beams
		\State \parbox[t]{\dimexpr\linewidth-\algorithmicindent}{Define sets $\mathcal{C}$ and $\mathcal{E}$ based on $\delta$\strut}
		\State \parbox[t]{\dimexpr\linewidth-\algorithmicindent}{Calculate required number of subbands ${n}'_k$ for \\each UE $k$ from (\ref{eq8})}
		\State \parbox[t]{\dimexpr\linewidth-\algorithmicindent}{Allocate ${n}'_k$ subbands to each UE $k$ in the order \\of increasing distance $\rightarrow$ $\Phi_{\theta}$ \strut}
		\State \parbox[t]{\dimexpr\linewidth-\algorithmicindent}{Calculate achievable data rate $R_k$ and \\$\Gamma_{\theta} = \sum_{k=1}^K\log(R_k)$\strut}
		\EndFor
		\State\parbox[t]{\dimexpr\linewidth-\algorithmicindent}{Determine $\theta^*= \arg\max_{\theta}\Gamma_{\theta}$ and allocate \\corresponding resources from $\Phi_{\theta^*}$\strut}  
		\EndFor
	\end{algorithmic}\label{algo1}
\end{algorithm}    

Given the circular structure of the beamsectors, we have 
\begin{equation*}
	\begin{cases}b_{v+1} = b_1 \in \omega_{m+1}, & v = V,\\b_{v-1} = b_V \in \omega_{m-1}, & v = 1 .\end{cases}   
\end{equation*}
and vice versa, as illustrated in Fig. \ref{sfig2}. In the next step, for each user, distances $a_k$ to the beam edge $b_v$ are calculated. Based on distances $a_k$ and predefined threshold $\delta$ we define $\mathcal{C} = \begin{Bmatrix} \text{UE}_k|a_k>\delta,1 \le k \le K \end{Bmatrix}$ and $\mathcal{E} = \begin{Bmatrix} \text{UE}_k|a_k \leq \delta,1 \le k \le K \end{Bmatrix}$, for all UEs with reported positions in $b_v$, $b_{v-1}$ and $b_{v+1}$. Next, the required number of subbands for each UE to satisfy its own rate requirement is determined as follows:
\begin{equation}\label{eq8}
	\left\{\begin{matrix}
	{n}'_k = \left \lceil \frac{R_{min}^{\mathcal{C}}}{W\log_2(1+{\gamma}'_{k,m,n})}  \right \rceil, & a_k > \delta, 
	\vspace{3pt} \\ 
	{n}'_k = \left \lceil \frac{R_{min}^{\mathcal{E}}}{W\log_2(1+{\gamma}'_{k,m,n})}  \right \rceil, & a_k \leq \delta, 
	\end{matrix}\right.
\end{equation}
where ${\gamma}'_{k,m,n}$ was given by \eqref{eq22} in Section \ref{ss2_A}.
Then, the required amount of subbands are allocated to each UE in the order of increasing distance, until no more subbands are available. After allocation, the proportional fairness metric $\Gamma_{\theta}$ and corresponding allocation matrix $\Phi_{\theta}$ are determined. Iterating over the possible values of beamwidth $\theta$, our algorithm finally selects the value of $\theta$ that maximizes proportional fairness.

\section{Performance Evaluation}\label{s5}
To evaluate the performance of the proposed algorithm, we consider a circular cell of radius 100m, with the BS at the center. The BS is equipped with six analog beamformers, each covering a sector of 60$^o$. The path-loss model is as in \cite{b4},
\begin{equation}
	PL[dB] = 98.4 + 20 \hspace{2pt} \log_{10}\hspace{2pt}f + 10 \hspace{2pt} \alpha \hspace{2pt} \log_{10} \hspace{2pt} R ,
\end{equation}
where $f$ [GHz] is the carrier frequency and $R$ [km] is the distance between transmitter and receiver. Moreover, we take into account small scale Rayleigh fading. A full buffer traffic model is assumed. The remaining system parameters are listed in Table \ref{tab1}, where QoS requirements for center and edge UEs are given as 2 Gbps and 1 Gbps, respectively. The beamwidth is adapted in each sector and every scheduling frame among six possible candidates. We consider two uncertainty levels $\beta$ and three threshold values $\delta$ for constraint (\ref{eq4e}). 
\begin{table}[t]
	\caption{Simulation Parameters}
	\begin{center}
		\normalsize	
		\begin{tabular}{l l}
			\hline
			\textbf{System Parameters}              & \textbf{Value}                               \\
			\hline
			\hline
			Coverage area                           & $\pi(100m)^2$                                \\ 
			System bandwidth                        & 1 GHz                                        \\ 
			Number of subbands $N$                  & 8                                            \\
			Number of transmitted beams $M$         & 6                                            \\ 
			Carrier frequency                       & 60 GHz                                       \\
			Noise power $N_0$                       & -174 dBm/Hz                                  \\
			Transmit power $P$                      & 30 dBm/sector                                  \\ 
			Candidate beamwidths $\theta$           & $3^o$, $5^o$, $10^o$, $15^o$, $20^o$, $30^o$ \\
			Pathloss exponent $\alpha$              & 2                                            \\ 
			QoS requirements $R_{min}^{\mathcal{C}}$, $R_{min}^{\mathcal{E}}$ & 2 Gbps, 1 Gbps                                        \\ [0.5ex]
			\hline
		\end{tabular}
		\label{tab1}
	\end{center}
	\vspace{-1.5em}
\end{table}

\begin{figure}
	\centering
	\includegraphics[width=0.99\linewidth]{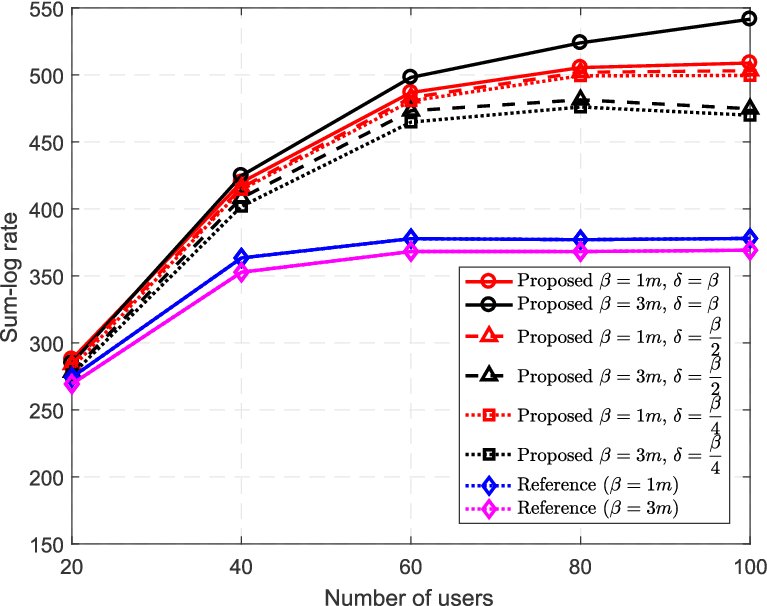}
	\caption{Proportional fairness performance of proposed and reference algorithms}
	\vspace{-1.0em}
	\label{fig4}
\end{figure}

\begin{figure}
	\centering
	\includegraphics[width=0.99\linewidth]{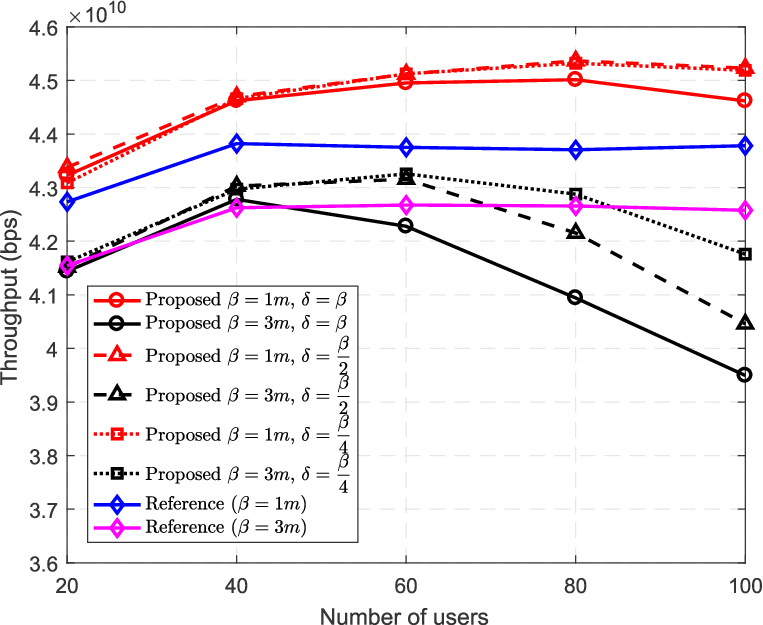}
	\caption{System throughput performance of proposed and reference algorithms}
	\vspace{-2.0em}
	\label{fig5}
\end{figure}

Fig. \ref{fig4} illustrates the average proportional fairness performance over 50000 frames of proposed and reference algorithms, for varying numbers of UEs. It can be observed that by considering position uncertainty, the proposed algorithm outperforms the reference algorithm where edge UEs are not protected. In particular, adjusting threshold $\delta$ such that $\delta = \beta$, provides the best performance compared to setting $\delta$ to lower levels. This shows the effectiveness of our algorithm in taking care of user position uncertainties. However, Fig. \ref{fig5} shows that the system throughput can suffer by setting the maximum threshold in the case of high uncertainty $\beta=3m$, since this leads to higher protection of edge users. This stresses the importance of selecting an optimized threshold depending on the uncertainty level. We observe that unlike in the case of proportional fairness, setting smaller thresholds for higher uncertainty levels enhances the system throughput, since the number of edge users is decreased. Nevertheless, for $\beta=1m$, as well as $\beta=3m$ with $\delta=\frac{\beta}{2}$ up to 80 users, the proposed scheme enables to simultaneously enhance the proportional fairness and throughput performances of reference schemes.

\begin{figure}
	\centering
	\includegraphics[width=0.98\linewidth]{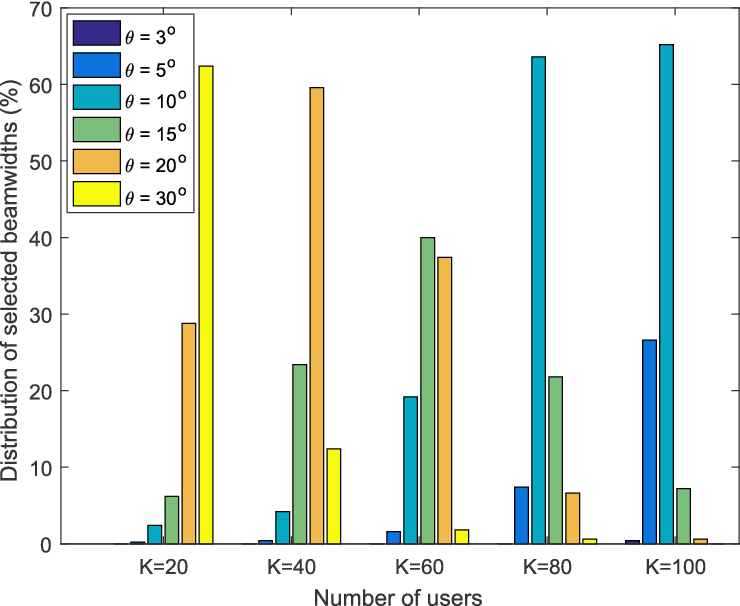}
	\caption{Distribution of selected beamwidths, depending on the number of UEs, ($\beta = 3m$, $\delta = \beta$)}
	\vspace{-1.5em}
	\label{fig6}
\end{figure}

Next, Fig. \ref{fig6} shows the distribution of beamwidths $\theta$ selected by our algorithm, for $\beta=3m$ and for different numbers of users. Clearly, the proposed algorithm adaptively selects all beamwidth values, while smaller values have a higher probability to be selected as the number of UEs $K$ increases. With larger $K$, the number of users within coverage of smaller beamwidths increases.  Hence, this enables smaller beamwidths to provide better throughput and higher fairness as well. Subsequently, the system proportional fairness is improved by selecting smaller beamwidths for larger $K$. From the figure, the beamwidth with maximum selection rate for each $K$ are $\theta=30^o, 20^o, 15^o, 10^o, 10^o$, respectively. Thus, by adequately selecting such intermediate values of $\theta$, the proposed algorithm enables to provide an efficient trade-off between system performance and robustness against position uncertainties and imperfect CSIs.

\section{Conclusion}
We proposed a fairness-aware joint adaptive beam-frequency allocation scheme for mmWave based mobile communication systems. In particular, the proposed method provides an efficient beamwidth adaptation to cope with the effects of user position uncertainties and imperfect CSIs. Simulation results show that, our algorithm largely enhances the system proportional fairness compared to the reference algorithm, even for large uncertainty levels. In the future work, we will provide a strategy to optimize over the edge users' threshold parameter as well, and extend the proposed scheme by including power allocation optimization and user mobility models. 

\end{document}